\documentclass[a4paper, 11pt]{article}

\usepackage{graphicx}
\usepackage{pdfpages}
\usepackage{color}
\usepackage[colorlinks=true,urlcolor=blue,citecolor=black]{hyperref}
\usepackage{url}
\usepackage[font=footnotesize,labelfont=bf]{caption}
\usepackage{float}
\usepackage{listings}
\usepackage{amsmath}

\usepackage[margin=1.25in]{geometry}

\title{\LARGE A Family of Multi-Asset Automated Market Makers}
\author{
    Eric Forgy\\
    \footnotesize\href{mailto:eric@cavalre.com}
    {\nolinkurl{eric@cavalre.com}}
	\and
    Leo Lau\\
    \footnotesize\href{mailto:leo@btx.capital}
    {\nolinkurl{leo@btx.capital}}
}   

\date{November 15, 2021}

\begin{document} 
\maketitle
\begin{abstract}
    We present a family of multi-asset automated market makers whose liquidity curves are derived from the financial principles of self financing transactions and rebalancing. The constant product market maker emerges as a special case.
\end{abstract}


\section{Introduction}

To fix notation, we consider $n$ asset tokens in a pool with $P_i^t$ denoting the price of token $i$ at time $t$ and $\alpha_i^t$ denoting the number of token $i$ in the pool at time $t$ so that
\[V_i^t = \alpha_i^t P_i^t\]
is the value of token $i$ at time $t$. In addition, we require the existence of a token specific to each pool that acts both as a liability and a numeraire to the other asset tokens so that all asset tokens are priced in terms of the pool token. We assign the index $i=0$ to the pool token so that the total value of the pool is given by
\[V^t = \sum_{i=0}^{n} \alpha_i^t P_i^t = \alpha_0^t + \sum_{i=1}^{n} \alpha_i^t P_i^t,\]
where we used the fact that since the pool token is the numeraire, its price (in terms of itself) is always $P_0^t = 1$. Furthermore, we require the total value of the pool to be zero so that the pool token is a liability with value given by
\[\alpha_0^t = -\sum_{i=1}^{n} \alpha_i^t P_i^t.\]
From this point, the derivation of the new liquidity curves follows from assuming:
\begin{itemize}
    \item All transactions are self financing.
    \item The pool has a specified rebalancing strategy.
\end{itemize}

\section{Self Financing}

For any variable $f_i^t$ relating to token $i$ at time $t$, let
\[\Delta f_i^t = f_i^t - f_i^{t-1}\]
denote its change from time $t-1$ to time $t$ and let
\[E_k f_i^t = k f_i^t + (1-k) f_i^{t-1}\]
denote a weighted average of the value between time $t-1$ and time $t$ with weight $0\leq k\leq 1$.
A discrete product rule can be written as
\[\Delta\left(f_i^t g_i^t\right) = \left(\Delta f_i^t\right) \left(E_k g_i^t\right) + \left(E_{1-k} f_i^t\right) \left(\Delta g_i^t\right).\]
The right-hand side of the above is independent of the choice of $k$, which merely specifies different decompositions of the product rule.

With this discrete product rule, the change in value of a token in the pool can then be expressed as
\begin{align*}
\Delta V_i^t 
&= \Delta\left(\alpha_i^t P_i^t\right) \\
&= \left(\Delta \alpha_i^t\right) \left(E_k P_i^t\right) + \left(E_{1-k}\alpha_i^t\right) \left(\Delta P_i^t\right).
\end{align*}
This can be decomposed into a change of value due to trading activity
\[\Delta V_{i,\text{trading}}^t\left(k\right) = \left(\Delta\alpha_i^t\right) \left(E_k P_i^t\right)\]
and a change in value due to price movements
\[\Delta V_{i,\text{price}}^t\left(k\right) = \left(E_{1-k} \alpha_i^t\right) \left(\Delta P_i^t\right).\]
A pool of tokens is self financing if the total change in value due to trading activity is zero, i.e.
\begin{align*}
\Delta V_\text{trading}^t\left(k\right)
&= \sum_{i=0}^n \Delta V_{i,\text{trading}}^t\left(k\right) \\
&= \sum_{i=0}^n \left(\Delta\alpha_i^t\right) \left(E_k P_i^t\right) \\
&= \Delta\alpha_0^t + \sum_{i=1}^n \left(\Delta\alpha_i^t\right) \left(E_k P_i^t\right) \\
&= 0
\end{align*}
so that
\begin{equation}
\label{eq:self-finance}
\alpha_0^t = \alpha_0^{t-1} -\sum_{i=1}^n \left(\Delta\alpha_i^t\right) \left(E_k P_i^t\right).
\end{equation}
It follows that the total value due to market movement is also zero because we assumed the total value of the pool is always zero, i.e.
\[\Delta V^t = \Delta V^t_\text{trading}\left(k\right) + \Delta V^t_\text{price}\left(k\right) = 0\]
so that
\begin{equation*}
    \sum_{i=1}^n \left(E_{1-k} \alpha_i^t\right) \left(\Delta P_i^t\right) = 0
\end{equation*}
where the sum starts at $i=1$ because $P_0^t=1$ for all $t$ so that $\Delta P_0^t = 0.$

\section{Rebalancing Strategies}

The final ingredient needed to derive the liquidity curve is to require a specified rebalancing strategy for the pool. The weight of an asset token $i$ in the pool at time $t$ is the ratio of its value to the total value of all asset tokens in the pool, i.e.
\[\omega_i^t = \frac{V_i^t}{\sum_{j=1}^n V_j^t}\]
for $j$ ranging over asset tokens $1$ to $n$ so that
\[\sum_{i=1}^{n} \omega_i^t = 1.\]
A rebalancing strategy is a specification of all weights at time $t$ given information no later than time $t-1$.
For a given rebalancing strategy, the price of a token  at time $t$ is given by
\begin{equation}
\label{eq:price}
P_i^t = -\omega_i^t \frac{\alpha_0^t}{\alpha_i^t}.
\end{equation}

\section{Liquidity Curves}

A liquidity curve is the foundation of an automated market maker (AMM) and represents a constraint among token number growths (and hence price changes) in an $n$-asset pool
\[g_0^t = \phi_k\left(g_1^t,\cdots,g_n^t\right),\]
where
\[g_i^t = \frac{\alpha_i^t}{\alpha_i^{t-1}}\]
is the token number growth factor for token $i$ from time $t-1$ to $t$. 

To derive the liquidity curves, we simply combine the self financing formula (\ref{eq:self-finance}) with the price formula (\ref{eq:price}) from the rebalancing strategy, i.e.
\begin{align*}
\alpha_0^t 
&= \alpha_0^{t-1} - \sum_{i=1}^n \left(\Delta \alpha_i^t\right) \left(E_k P_i^t\right) \\
&= \alpha_0^{t-1} - \sum_{i=1}^n \left(\Delta \alpha_i^t\right) \left[E_k\left(-\omega_i^t \frac{\alpha_0^t}{\alpha_i^t}\right)\right].
\end{align*}
The above can be rewritten as a one-parameter family of liquidity curves given by
\begin{equation}
    \label{eq:gen-curve}
    g_0^t = \phi_k\left(g_1^t,\cdots,g_n^t\right) = \frac{k + \left(1-k\right) \sum_{i=1}^n  \omega_i^{t-1} g_i^t}{\left(1-k\right) + k \sum_{i=1}^n\omega_i^t\left(g_i^t\right)^{-1}}.
\end{equation}
If we assume equal weights so that $\omega_i^t = 1/n$, then we have the liquidity curves
\begin{equation}
    \label{eq:equal-curve}
    g_0^t = \frac{n k + \left(1-k\right) \sum_{i=1}^n g_i^t}{n \left(1-k\right) + k \sum_{i=1}^n \left(g_i^t\right)^{-1}}.
\end{equation}

\section{Asset Swaps}

Consider an equally-weighted pool with $n$ asset tokens and we wish to swap the first two. It follows that $g_0^t = 1$ and the liquidity curve becomes
\begin{equation}
    \label{eq:equal-swap}
    \left(1-k\right)\left(g_1^t+g_2^t-2\right)=k\left(\frac{1}{g_1^t}+\frac{1}{g_2^t}-2\right),
\end{equation}
which is notably independent of $n.$ When $k=1/2$, the liquidity curve simplifies further to
\begin{equation*}
    g_1^t+g_2^t=\frac{1}{g_1^t}+\frac{1}{g_2^t},
\end{equation*}
which is equivalent to the familiar constant product market maker
\begin{equation}
    \label{eq:uniswap-curve}
    g_1^t g_2^t = 1 \implies \alpha_1^t \alpha_2^t = \alpha_1^{t-1} \alpha_2^{t-1}=\text{constant}
\end{equation}
as originally used by Uniswap \cite{uniswap2} for $n=2$ and Balancer \cite{balancer} for $n>2$.

\begin{figure}[htb]
\centering 
\includegraphics[width=.8\textwidth]{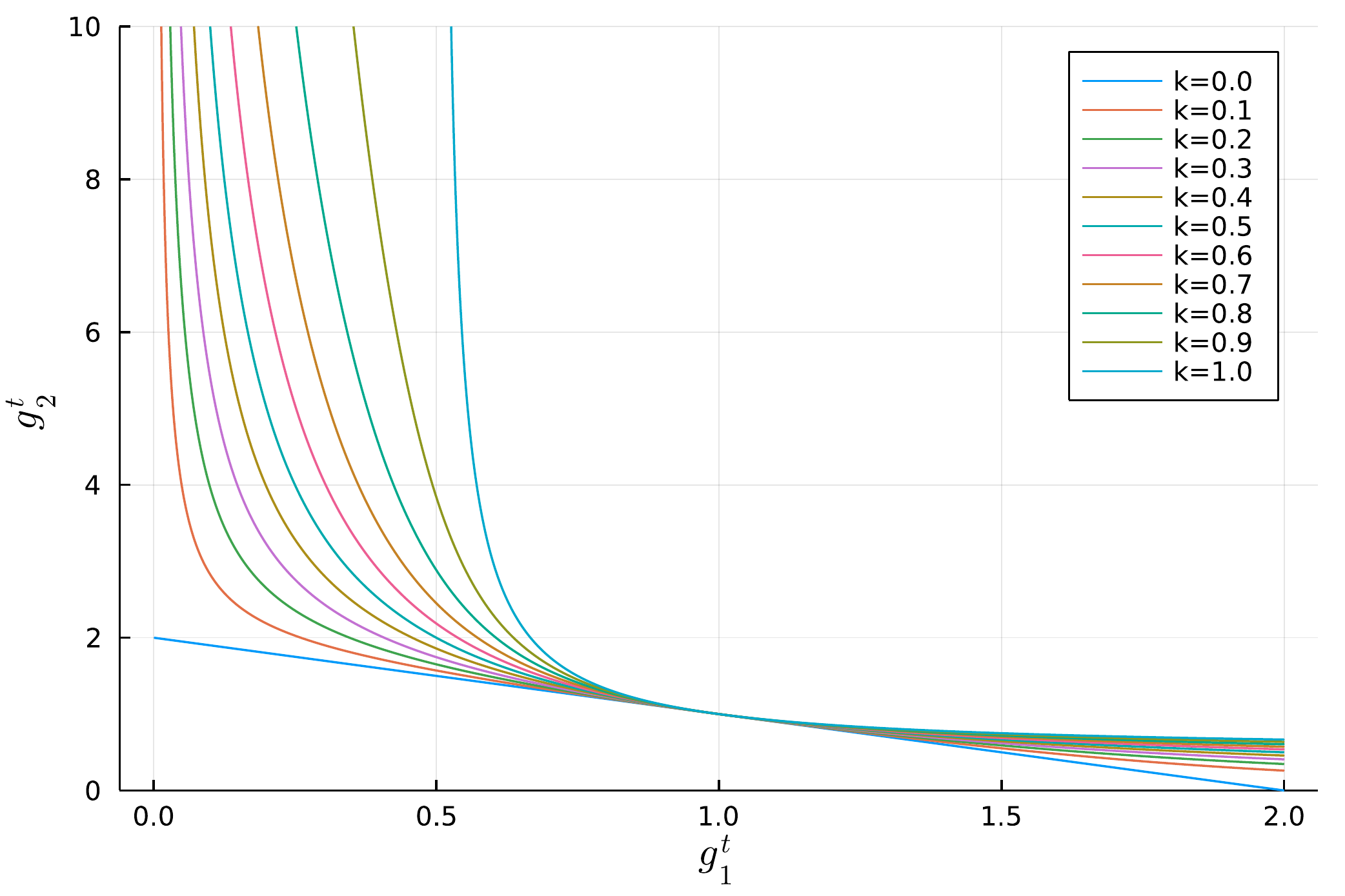}
\caption{\label{fig:swap} Liquidity curves for asset swaps.}
\end{figure}

Liquidity curves for various values of $k$ are illustrated in Figure \ref{fig:swap}. 
All curves are bounded below by $k=0$ with liquidity curve
\begin{equation}
    \label{eq:k=0 curve}
    g_1^t + g_2^t = 2
\end{equation}
and above by $k=1$ with liquidity curve
\begin{equation}
    \label{eq:k=1 curve}
    \frac{1}{g_1^t} + \frac{1}{g_2^t} = 2
\end{equation}
with the constant product curve $k=1/2$ in between.

\begin{figure}[htb]
\centering 
\includegraphics[width=.8\textwidth]{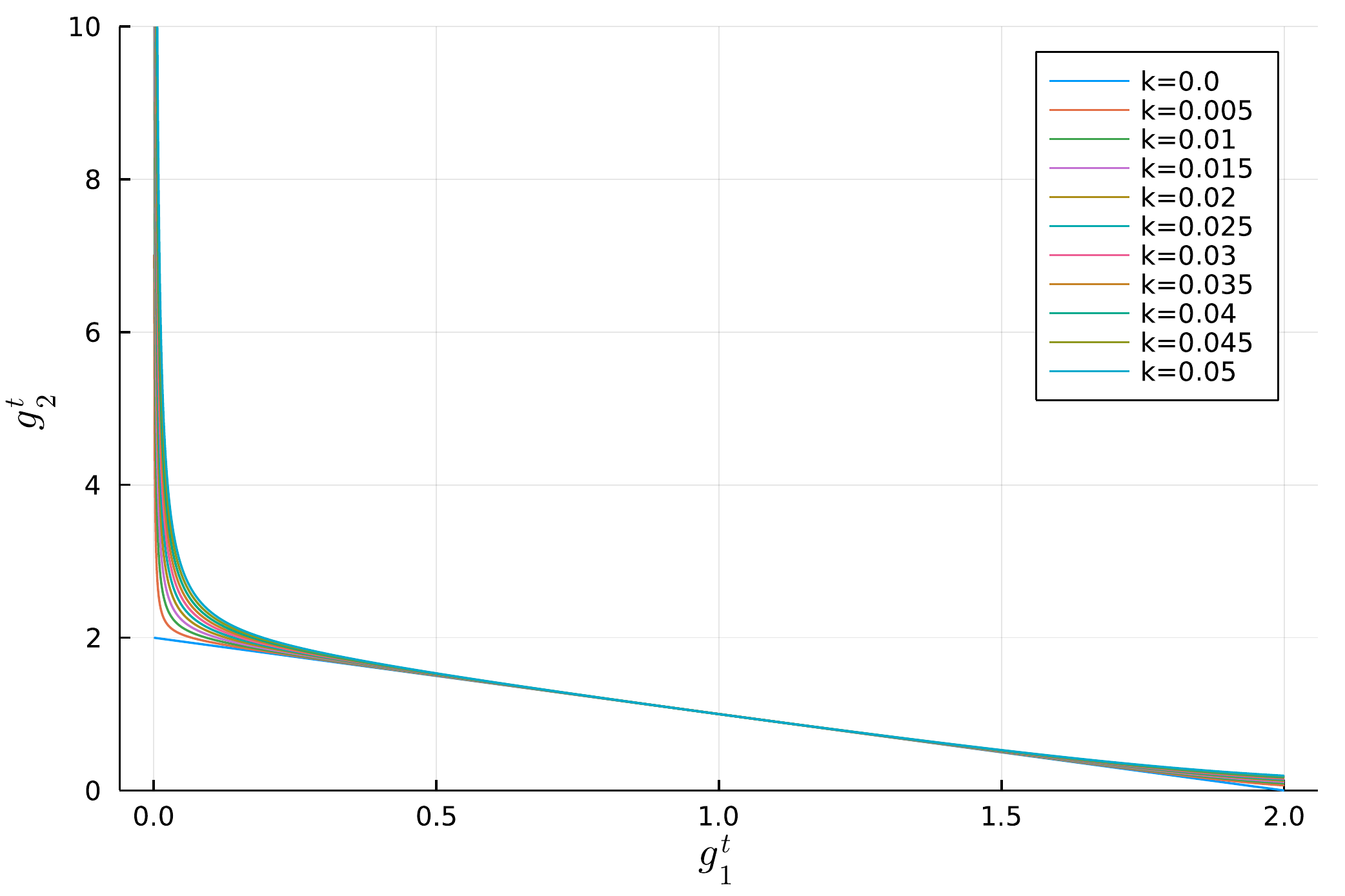}
\caption{\label{fig:swap0} Liquidity curves for asset swaps near $k=0.$}
\end{figure}

The lowest liquidity curve corresponding to $k=0$ does not produce a practical market maker since the constraint is linear even when all of one token is swapped out of the pool. However, for all $0< k \le 1,$ we get the desired nonlinearities and letting $k$ approach 0 shifts the nonlinearity closer to $g_1^t=0$ as illustrated in Figure \ref{fig:swap0}.
These liquidity curves near $k=0$ are most appropriate for swaps between liquid stablecoins referencing the same fiat currency.

\begin{figure}[htb]
\centering 
\includegraphics[width=.8\textwidth]{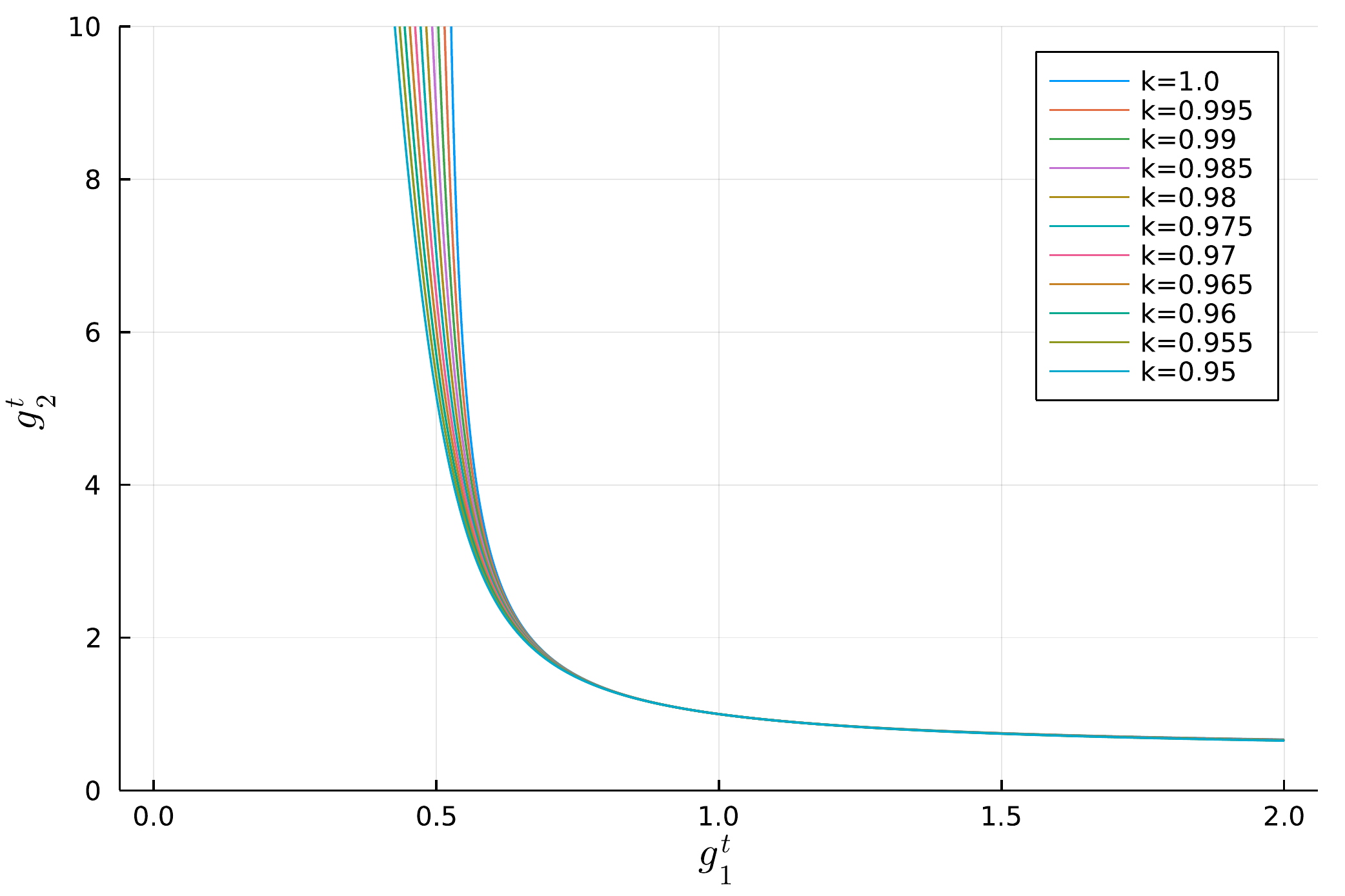}
\caption{\label{fig:swap1} Liquidity curves for asset swaps near $k=1.$}
\end{figure}

At the other extreme, when $k=1$ there is a discontinuity at $g_1^t = 1/2$  since
\begin{equation}
    \label{eq:discontinuity-2}
    \frac{1}{g_1^t} + \frac{1}{g_2^t} = 2\implies g_2^t = \frac{g_1^t}{2g_1^t-1}.
\end{equation}
This discontinuity represents a valid market maker, but with a restriction that no more than half the asset tokens in a pool can be swapped out in a single transaction. The liquidity curve for $k=1$ is a special case and for any other $0< k <1$, the liquidity curves all spike as $g_1^t$ approaches zero, i.e. as all of asset token 1 is removed from the pool in a single transaction. Liquidity curves near $k=1$ are illustrated in Figure \ref{fig:swap1}.
These market makers maximize the price impact of all trades and would be especially suitable for illiquid asset tokens, but remain suitable for any asset token especially when each transaction is small relative to the total size of the pool. 

The constant product market maker corresponding to $k=1/2$ is a special case of the general one-parameter family of automated market makers.

\section{Staking}

Staking is simply a swap that involves receiving pool tokens in exchange for one or more asset tokens. There are no pool tokens in the pool per se so when asset tokens are staked, pool tokens are minted. Conversely, when asset tokens are unstaked, pool tokens are burned.

\begin{figure}[htb]
\centering 
\includegraphics[width=.8\textwidth]{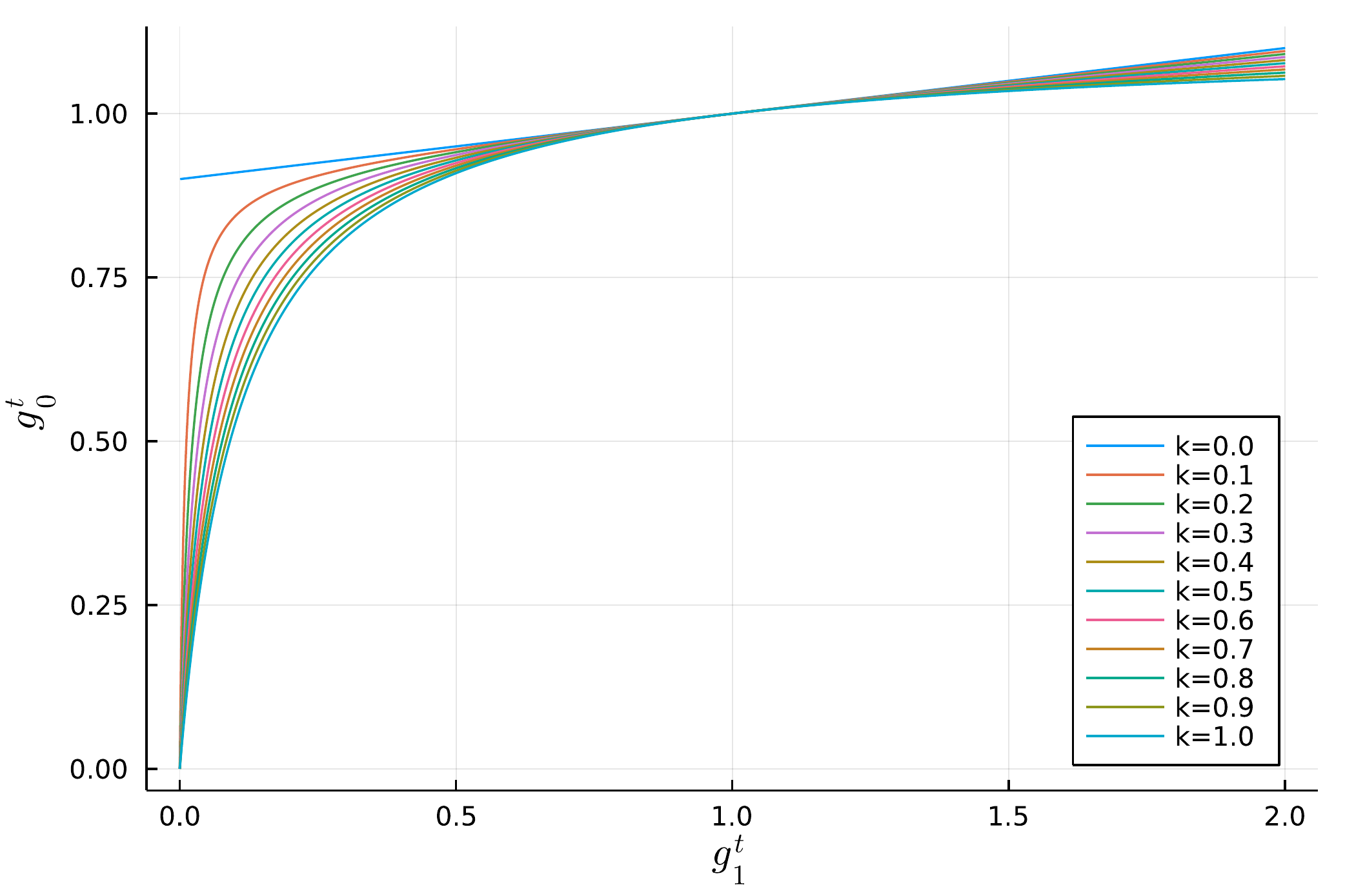}
\caption{\label{fig:stakek} Liquidity curves for single-asset staking with $n=10.$}
\end{figure}

\begin{figure}[htb]
\centering 
\includegraphics[width=.8\textwidth]{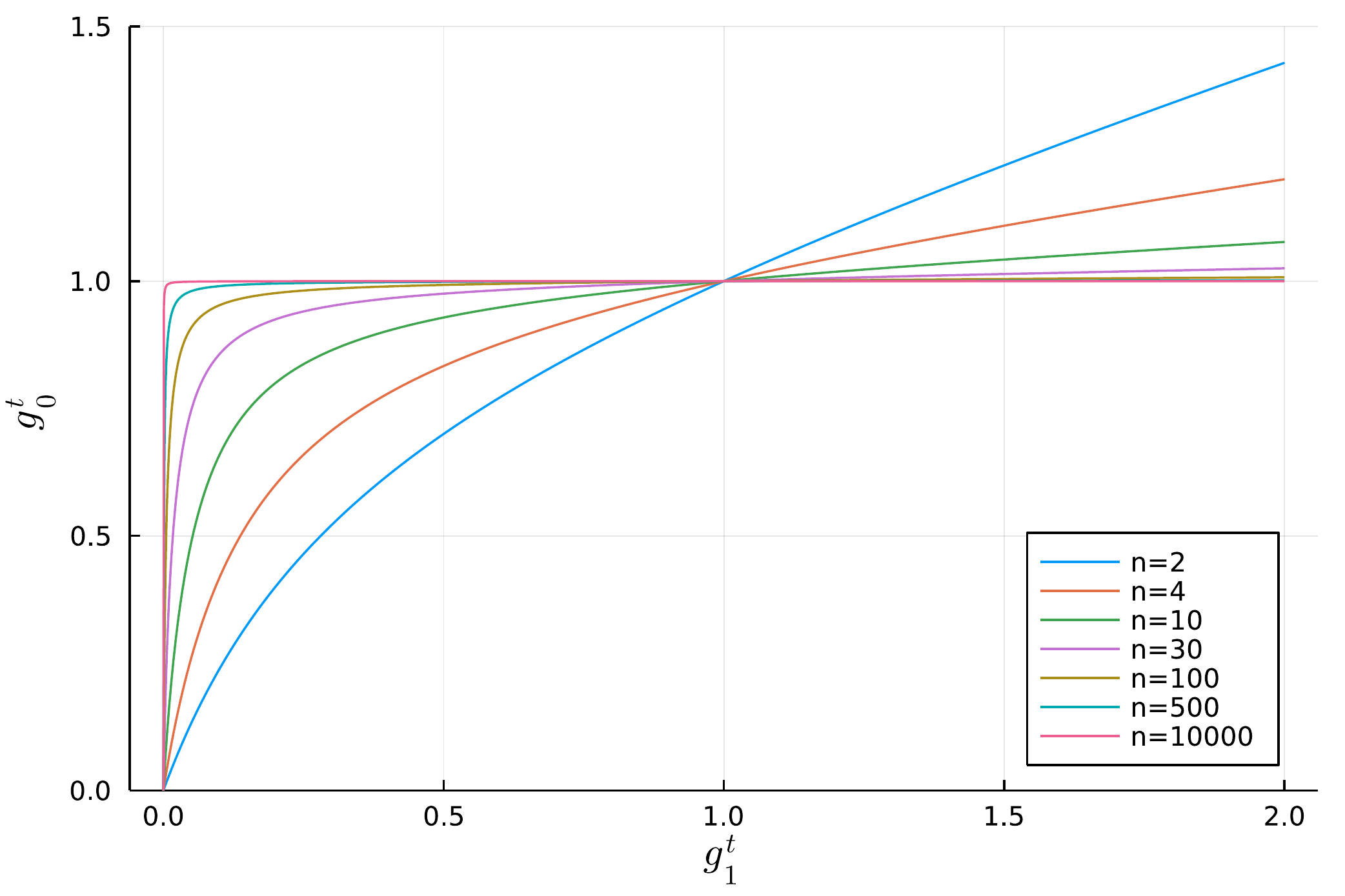}
\caption{\label{fig:staken} Liquidity curves for single-asset staking with $k=1/2.$}
\end{figure}

When asset tokens are staked in the same proportion, i.e. $g_i^t = g^t$, the liquidity curve collapses trivially to
\[g_0^t = g^t\]
and the pool token grows in the same proportion. When $n=2$ and $k=1/2,$ this corresponds to Uniswap and its cousins. However, when a single asset token is staked, the liquidity curve may be written as
\begin{equation}
    g_0^t = \frac{n + \left(1-k\right) \left(g_1^t-1\right)}{n + k \left[\left(g_1^t\right)^{-1}-1\right]}
\end{equation}
and the number of pool tokens minted depends on both the number of asset tokens in the pool and the value of $k$. Figure \ref{fig:stakek} illustrates the one-asset staking curve as a function of $k$ for a 10-asset pool.
Similarly, Figure \ref{fig:staken} illustrates the one-asset staking curve as a function of the number of asset tokens for $k=1/2.$
The case of $n=2$ and $k=1/2$ in Figure \ref{fig:staken} illustrates what a version of Uniswap would look like if it allowed one-sided staking. As the number of asset tokens in the pool increases, shared liquidity among asset tokens naturally mutes the price impact, i.e. flattens the liquidity curve, when staking and unstaking.



\section{Multi-Token Swaps}

In the preceding analysis, we considered multi-token swaps when all asset tokens are swapped in the same proportion and we considered simple swaps between any two tokens, where a two-token swap involving the pool token represents one-sided staking. However, the proposed model facilitates more general multi-token swaps involving any number of tokens in any proportions. The equally-weighted $n$-asset liquidity curve (\ref{eq:equal-curve}) contains $n+1$ degrees of freedom in the form of the token number growth factors $g_i^t$. If we specify any $n$ growth factors, we can use the liquidity curve to solve for the remaining growth factor. 

For example, if we wish to swap the first $m$ asset tokens in an $n$-asset pool so that $g_i^t = 1$ for $i>m$ and $g_0^t=1,$ the liquidity curve becomes
\begin{equation}
    \label{eq:equal-multiswap}
    \left(1-k\right)\left(-m+\sum_{i=1}^m g_i^t\right) = k \left(-m+\sum_{i=1}^m \frac{1}{g_i^t}\right)
\end{equation}
which is a generalization of (\ref{eq:equal-swap}) and, again, is notably independent of $n.$ To settle this multi-asset swap, we would need to first specify the growth factors for $m-1$ of the assets and use the liquidity curve (\ref{eq:equal-multiswap}) to solve for the remaining growth factor.

As a special case when $k=0$, the general equally-weighted $n$-asset liquidity curve (\ref{eq:equal-curve}) simplifies to
\[g_0^t = \frac{1}{n} \sum_{i=1}^n g_i^t,\]
i.e. the growth in pool tokens is the average growth of the asset tokens for a single transaction. If you wished to stake any number of asset tokens in any proportion, the liquidity curve tells you how many pool tokens you would receive. Similarly, when $k=1$, the liquidity curve becomes
\[\frac{1}{g_0^t} = \frac{1}{n} \sum_{i=1}^n \frac{1}{g_i^t}\]
and when $k=1/2,$ which is related to Uniswap for $n=2$ and Balancer for $n>2$, we have
\[g_0^t = \frac{n + \sum_{i=1}^n g_i^t}{n + \sum_{i=1}^n \left(g_i^t\right)^{-1}}.\]
In particular, when $k=1/2$ and $g_0^t=1$, we have
\[\sum_{i=1}^n g_i^t = \sum_{i=1}^n \left(g_i^t\right)^{-1},\]
which is valid for any arbitrary multi-asset swap, but swapping any two asset tokens $a$ and $b$ in the pool (like Uniswap / Balancer) results in 
\[g_a^t g_b^t = 1\implies \alpha_a^t \alpha_b^t = \text{constant}\]
as highlighted in (\ref{eq:uniswap-curve}) above.



\section{Conclusions and Future Work}

We presented a one-parameter family of liquidity curves in (\ref{eq:gen-curve})
derived from simple financial principles assuming all trades are self financing and that the pool imposes a rebalancing strategy. These liquidity curves then form the basis for a new one-parameter family of AMMs. 

When the asset token weights are equal, we saw that the liquidity curves were independent of the number of asset tokens in the pool for pure asset token swaps, but the shared liquidity among asset tokens reduces the price impact for swaps that involve the pool token when staking and unstaking.

We saw that liquidity curves near $k=0$ were largely insensitive to the size of the swap and are most suitable for swapping stablecoins referencing the same fiat currency. The liquidity curve for $k=1$ represents a special case with a restriction that no more than half of any asset token in the pool can be swapped out in a single transaction for a 2-asset swap.

The purpose of this report is simply to introduce the new family of AMMs. Subsequent reports will compare the proposed model to existing AMMs in the spirit of \cite{ammmath} and references therein. In particular, we plan to look at issues of capital efficiency, transaction fees and impermanent loss. 

In addition to general multi-token swaps, the model also accommodates batching of trades making it particularly amenable to implementation on layer 2 rollups such as zkSync \cite{zksync} or StarkNet \cite{starknet}.

Uniswap v3 \cite{uniswap3} introduces a number of innovations including concentrated liquidity and flexible fees. We plan to bring those innovations and others to this family of models as well and do not foresee any difficulties in doing so.

\section*{Acknowledgement}

The authors would like to thank Dan Robinson for helpful discussions during the early development of this work.

\bibliographystyle{unsrt}
\bibliography{amm}

\begin{thebibliography}{1}

\bibitem{uniswap2}
Hayden Adams, Noah Zinsmeister, and Dan Robinson.
\newblock {Uniswap v2 Core.}
\newblock \url{https://uniswap.org/whitepaper.pdf}, Mar 2020.

\bibitem{balancer}
Fernando Martinelli and Nikolai Mushegian.
\newblock {A non-custodial portfolio manager, liquidity provider, and price
  sensor.}
\newblock \url{https://balancer.fi/whitepaper.pdf}, Sep 2019.

\bibitem{ammmath}
Leo Lau and Guangwu Xie.
\newblock {A Mathematical View of Automated Market Maker (AMM) Algorithms and
  Its Future}.
\newblock
  \url{https://medium.com/anchordao-lab/automated-market-maker-amm-algorithms-and-its-future-f2d5e6cc624a},
  Sep 2021.

\bibitem{zksync}
{zkSync}.
\newblock \url{https://zksync.io/}.

\bibitem{starknet}
{StarkNet}.
\newblock \url{https://starkware.co/starknet/}.

\bibitem{uniswap3}
Hayden Adams, Noah Zinsmeister, Moody Salem, River Keefer, and Dan Robinson.
\newblock {Uniswap v3 Core.}
\newblock \url{https://uniswap.org/whitepaper-v3.pdf}, Mar 2021.

\end{thebibliography}

\end{document}